\documentclass[authoryear,preprint,review,12pt]{elsarticle}

\usepackage{amssymb}
\usepackage{amsmath}
\usepackage{amsthm}
\usepackage{bm}
\usepackage{float}
\usepackage{subfig}
\usepackage{booktabs}   
\usepackage{tabularx}      
\usepackage{threeparttable}
\usepackage[colorlinks=true,linkcolor=blue,citecolor=blue,urlcolor=blue]{hyperref}

\journal{Journal of Theoretical Biology}

\begin{document}

\begin{frontmatter}



\title{mRNA-protein assembly reduces fluctuations in a system with bursty transcription} 


\author[HKU]{Xinke Lyu} 
\ead{u3597334@connect.hku.hk}
\author[FlatIron]{Alex Mayer}
\ead{ajmayer95@yahoo.com}
\author[Duke,UNC]{Grace McLaughlin}
\ead{mclaughlingrace@gmail.com}
\author[Duke]{Amy Gladfelter}
\ead{amy.gladfelter@duke.edu}
\author[UCLA,CDM]{Marcus Roper\corref{cor1}}
\ead{mroper@math.ucla.edu}
\cortext[cor1]{Corresponding author}
\affiliation[HKU]{organization={ Department of Mathematics, The University of Hong Kong},
            city={Hong Kong},
            country={China}}

\affiliation[UCLA]{organization={ Department of Mathematics, UCLA},
            city={Los Angeles},
            postcode={90095}, 
            state={California},
            country={USA}}
\affiliation[CDM]{organization={ Department of Computational Medicine, UCLA},
            city={Los Angeles},
            postcode={90095}, 
            state={California},
            country={USA}}
\affiliation[FlatIron]
{organization = {Center for Computational Biology, Flatiron Institute}, 
city={New York}, 
state= {New York},
postcode = {10010}, 
country = {USA}}
\affiliation[Duke]
{organization = {Dept. of Cell Biology, Duke University School of Medicine},
city = {Durham},
state = {North Carolina},
postcode = {27708},
country = {USA}}

\affiliation[UNC]
{organization = {Dept. of Biology, University of North Carolina at Chapel Hill},
city = {Chapel Hill},
state = {North Carolina},
postcode = {27599},
country = {USA}}            
\begin{abstract}
mRNA-protein assemblies play a fundamental role in forming membraneless compartments within cells, whose functions may include activating, inhibiting, and localizing reactions. Recruitment of proteins into droplets can diminish cell to cell variability in protein abundance. However, the extent to which mRNA-protein assemblies may also buffer noise arising from transcription is not understood. Complicating study of this question is that models of kinetics typically treat this as a phase separation process, when mRNA-protein assemblies can contain as few as 2 mRNA transcripts, far below the thermodynamic thresholds for phase separation. Here, through stochastic simulations and asymptotic analysis, we quantify noise suppression by mRNA-protein assemblies as a function of gene expression kinetic parameters, and show that assemblies formed from just a handful of mRNAs effectively regulate transcript abundances and suppress fluctuations. We place particular emphasis on how this mechanism can facilitate regulated transcription by reducing noise even in the context of infrequent bursts of transcription. We investigate two biologically relevant models in which mRNA assembly acts to either ``buffer'' noise by storing mRNA in inert droplets, or ``filter'' assembled mRNAs by accelerating their decay, and quantify expression noise as a function of kinetic parameters. In either case, the most controlled expression occurs when bursts produce mRNAs close to the assembly threshold, which we find to be broadly consistent with observations of an RNP-droplet forming cyclin in multinucleate \textit{Ashbya gossypii} cells.
\end{abstract}

\begin{graphicalabstract}
\includegraphics[width=1\textwidth]{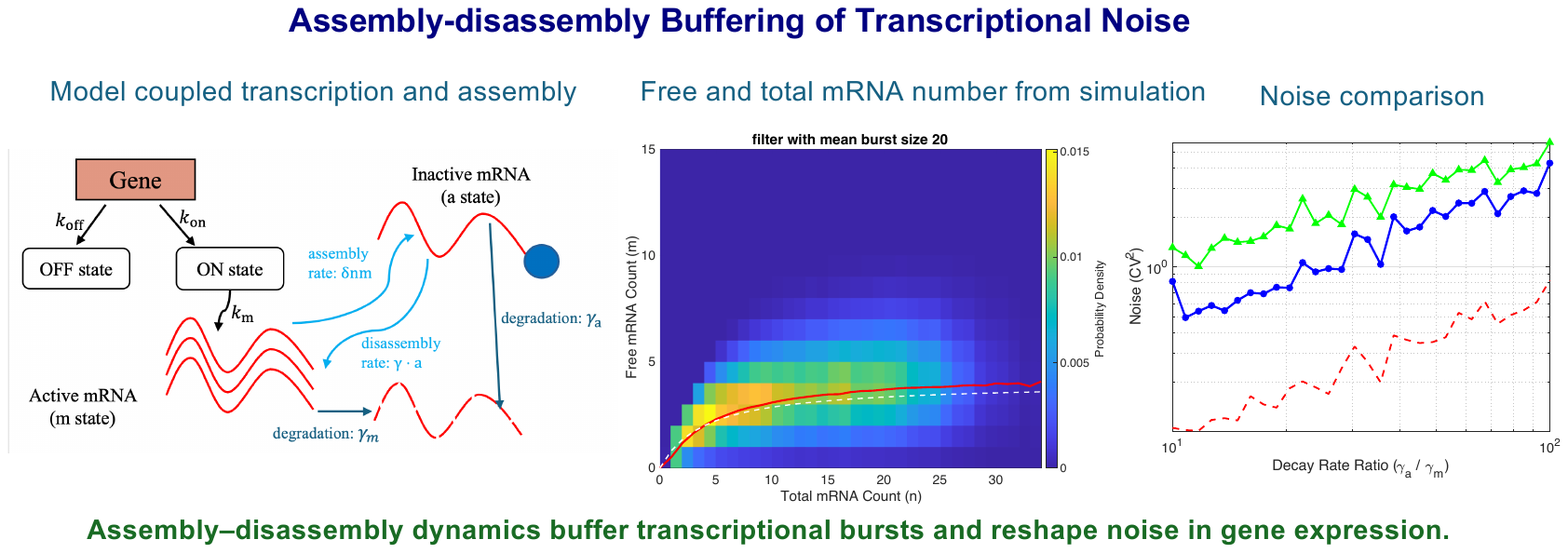}
\end{graphicalabstract}

\begin{highlights}
\item Assembly-disassembly dynamics reduce noise in gene expression.
\item Effective buffering mechanism even at low mRNA copy numbers.
\item Study fluctuation in the coupled model of transcription and assembly.
\end{highlights}

\begin{keyword}
mRNA-protein assembly \sep stochastic modeling \sep gene expression variability


\end{keyword}

\end{frontmatter}


\section*{Introduction}
Gene expression is a noisy process, ensuring that even genetically identical cells that receive identical cues from their environment may exhibit a range of protein copy numbers. In some cases, cell-to-cell variability might be useful: for example, in \textit{Bacillus subtilis}, the excitable competence network is triggered by stochastic pulses of gene expression \citep{maheshri2007living, raser2005noise}. However, in stable environments, cells generally benefit from consistent expression of proteins, and high expression noise may impair cellular function \citep{charlebois2015effect, wang2011impact, bahar2006increased}. Low mRNA copy numbers are a major contributor to protein copy noise since translation to proteins amplifies small absolute variations in mRNA copy number \citep{buccitelli2020mrnas}. Yet, mRNA copy numbers are often low - at 10 mRNAs or fewer per cell, across the majority of the genome in \textit{Saccharomyces cerevisiae} \citep{miura2008absolute}, and typically transcription rates are much smaller than translation rates across the genomes of yeast, mice, humans and \textit{E. coli} in their fast-growing phases \citep{hausser2019central}. \cite{hausser2019central} argue that energetics constrain transcription rates; the total energetic cost of translation is invariant if protein copy numbers are held fixed, but the energetic burden of transcription, although relatively smaller, is reduced if mRNA copy numbers are kept small. Lessening transcriptional interference may endow even greater benefits: the negative interactions of different transcriptional activities due, for example, to elongating RNA polymerases obstructing each other, repressors bound to one operon overlapping with a second operon \citep{shearwin2005transcriptional}, or from genome conformational changes that expose one operon, but mask another \citep{kim2019mechanisms}, reward cells for minimizing the number of shifts from expressing one gene to a different one.

In eukaryotes, the primary mode of transcription is through bursting, where mRNAs are synthesized in pulses \citep{nicolas2017shapes}. These pulses are often modeled as Poisson point processes, though evidence suggests that the arrival process may be more complex \citep{kumar2015transcriptional}.
Bursts are thought to arise as a result of the reversible interactions of the pre-initiation complex (PIC) with a gene's promoter. 
When the PIC is bound, the RNA polymerase's affinity for binding to the transcription start site is dramatically 
increased, and multiple transcription events can occur while the gene is in this ``ON" state \citep{boeger2015structural}. 
Since chromatin remodeling is required for gene activation, 
there is a fundamental cost to fast switching between the ``ON'' state and ``OFF'' state \citep{huang2015fundamental}. Efficiency can thus be achieved by minimizing the frequency of switching into the ``ON'' state and meeting the required mean mRNA abundance via intense bursts. However, bursty transcription introduces additional noise into expression. In particular, infrequent transitions into the ``ON'' state combined with intense bursts will drastically increase mRNA variability, particularly in systems with small mean mRNA transcription numbers, potentially resulting in large steady-state fluctuations in protein levels.

For many genes, protein abundance must be tightly regulated for proper function of the cell \citep{barkai2000circadian, lehner2008selection, wang2011impact}, and there exist many regulatory mechanisms that can reduce fluctuations in protein abundance \citep{tan2021quantitative, singh2011negative}. In particular, negative feedback has been observed at each level of gene expression, and previous mathematical models have quantified the noise control effect of different feedback mechanisms \citep{singh2011negative}. Layers of feedback can theoretically drive expression noise below Poisson levels, at the cost of introducing substantial deficits in the cell's energy economy \citep{stoeger2016passive}. For example, \cite{lestas2010fundamental} find that in systems with nonlinear real-time feedback control, signal molecules must be synthesized at rates far exceeding those of the target molecule in order to meaningfully suppress noise. By contrast, cellular compartmentalization may be an energetically efficient method for buffering expression noise if the proteins that form the compartments are long-lived  \citep{stoeger2016passive}. In particular, phase separation has been theorized to play a role in post-translational regulation of genes. \cite{klosin2020phase} provide theoretical and experimental evidence that concentration-dependent phase separation of proteins can drive protein fluctuations below the minimum Poisson noise limit of the network. They implicate phase separation in acting as a buffer that maintains the free molecules at the phase separation threshold. Deviri and Safran  \citep{deviri2021physical} extended this theory to multi-component phase separation, and derived criteria on their phase diagrams under which concentration buffering may occur. They hypothesized that in genes that are sensitive to noise selective pressures may act to optimize concentration buffering.

One such class of multi-component condensates is ribonucleoprotein (RNP) granules, which form as a result of multivalent interactions between mRNAs and RNA-binding proteins. RNA-binding proteins often contain intrinsically disordered domains, which promote RNP granule assembly 
and contribute to their dynamic properties \citep{tauber2020mechanisms}. RNP granules can subcompartmentalize the cytosol for regulated colocalization or segregation of interacting proteins and RNAs \citep{langdon2018new}. Notably, 
mRNAs that are sequestered into droplet phases may, in some cases, be inaccessible to translation  \citep{tsang2019phosphoregulated}, reducing effective mRNA copy numbers within the cell. Although an initial expectation may be that reducing free mRNA copy numbers would increase the noise in protein numbers, here we analyze how, by reducing fluctuations in mRNA copy numbers, assemblies may paradoxically reduce the noise of gene expression. Our modeling is differentiated from previous analyses of mRNA-protein and protein-protein droplet formation, which analyze it as a phase separation phenomenon. Although phase separation models may illuminate assemblies formed by abundant macromolecules, assemblies can form from very small numbers of mRNAs, invalidating thermodynamic approximations.

The motivation for us to study the dynamics in cases where mRNA abundances are not limitingly large is that small numbers of mRNA are the most vulnerable to variability, and our own data on \textit{CLN3} mRNA (a G1 cyclin transcript, translated into the cyclin Cln3) droplets stochastically forming within the fungus \textit{Ashbya gossypii} cells indicate that the threshold abundances at which mRNAs start to assemble into droplets are very small: as few as 2-3 mRNAs per assembly.

In this paper, we analyze several biophysical scenarios in which mRNAs segregate into assemblies. The central and unifying assumption in all our models will be that mRNAs in the droplet phase are primarily translationally inert, 
so that only ``free" mRNAs are accessible to ribosomes. We first approximate the dynamics of bursting mRNA state expression systems by a random-telegraph process, providing baseline results with which to compare systems in which assembly is incorporated. We then discuss several models that incorporate nonlinear transition rates to emulate the physics of assembly. The models will chiefly be differentiated by the way in which the assembly affects mRNA stability. Presence of RNPs may either decrease or increase mRNA lifetimes, even when homologous RNA/protein pairs are expressed in closely related species. For example, the mRNA \textit{CLN3} condenses into droplets with a protein partner Whi3. When Whi3 is deleted in \textit{S. cerevisiae} cells, \textit{CLN3} lifetimes increase \citep{cai2013effects}, suggesting that condensed mRNAs turn over more quickly than free mRNAs in the cytosol. Conversely, in the genetically similar filamentous fungus \textit{Ashbya gossypii}, assembly increases \textit{CLN3} lifetimes \citep{lee2013protein}. 
 To address both of these functions of RNP bodies, we separately consider limiting cases where mRNA decay primarily occurs in the cytoplasm or when it occurs in the assemblies. In both models, we run stochastic simulations to determine how these mechanisms influence noise in gene expression, with particular focus on gene networks with infrequent, bursty transcription. Specifically, two models are proposed: a ``buffer model,'' where assemblies protect mRNA from degradation, and a ``filter model,'' where assemblies accelerate mRNA degradation. To support our theoretical findings, we perform analysis on existing smFISH data on the distribution of \textit{CLN3} mRNA transcripts within cells of the model filamentous fungus, \textit{Ashbya gossypii}.

While the capacity of phase-separated macromolecules to buffer noise is increasingly recognized \citep{klosin2020phase}, the main focus of our study is to investigate how assembly with proteins may drive noise reduction when mRNA copy numbers are small, and thermodynamic approximations can not be used, a scenario that is applicable to molecules present in low or very low numbers, such as mRNAs. 
In this study, we develop and analyze a model of mRNA assembly and disassembly. To investigate how assembly modulates noise from transcriptional bursting, we embed this mechanism into a fully coupled stochastic model. This framework dynamically links our downstream assembly model with the upstream process of gene expression, where stochastic bursting dynamics are reproduced using the Gillespie algorithm. By systematically varying the rates of transcriptional bursts and the effects of assemblies on mRNA stability, we analyze the system's steady-state behavior and noise level (measured by $CV^2$). Moreover, we demonstrate that the noise-reducing capability of assembly is still robust when governed by these inherently stochastic and nonlinear processes, facilitated by a small number of mRNAs.

\section*{Coupled transcription and assembly model}
\begin{figure}[htbp]
    \centering
    \includegraphics[width=1\linewidth]{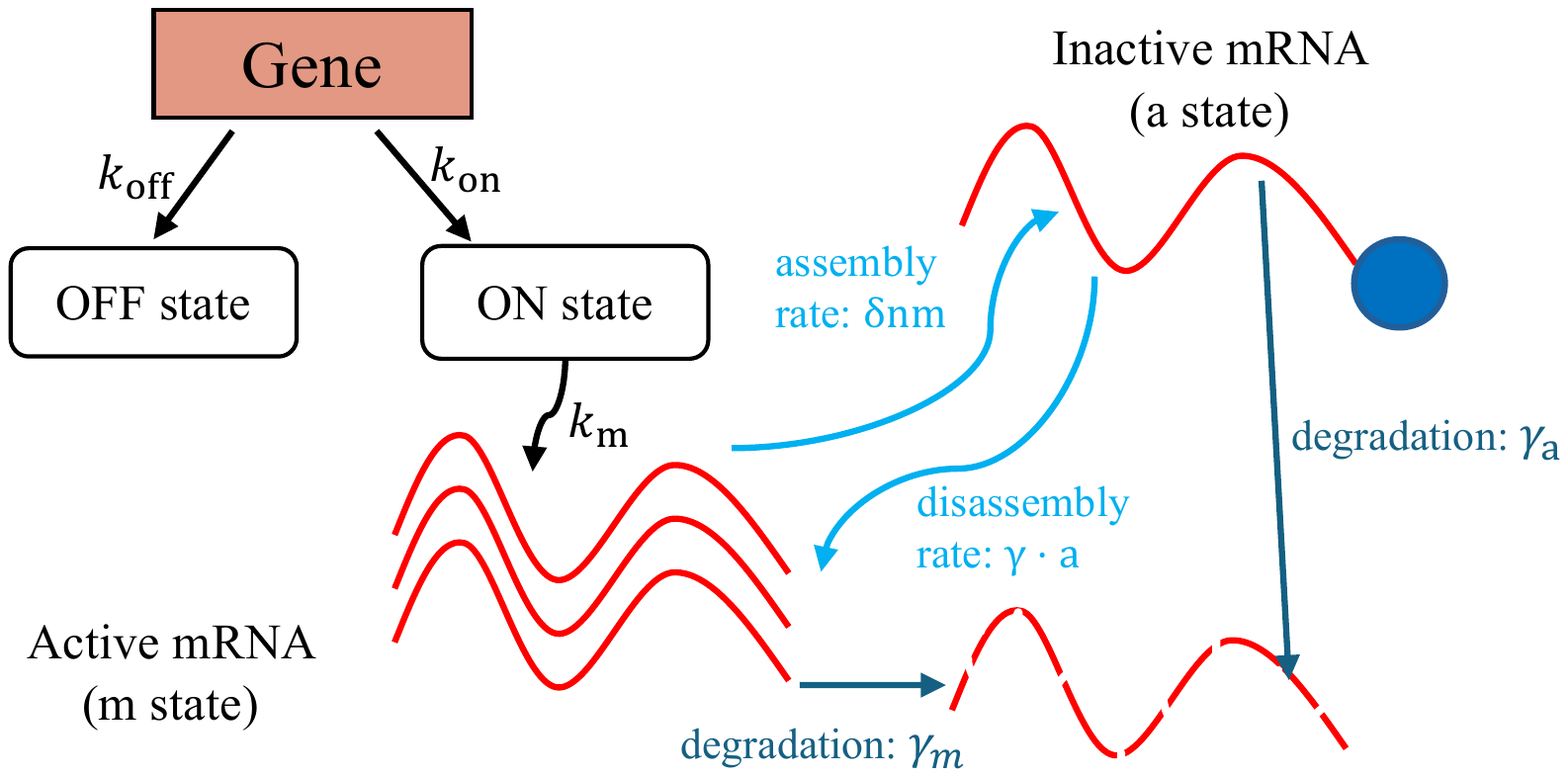}
    \caption{Unified stochastic model for transcription of new mRNAs in bursts, their degradation, assembly, and disassembly}
    \label{fig:1}
\end{figure}

The full model (\autoref{fig:1}) integrates transcriptional bursting, mRNA degradation, mRNA assembly, and disassembly in a single Gillespie stochastic simulation \citep{boeger2015structural}.\\
It tracks the following variables: $g\in \{0,1\}$ represents gene transcription OFF/ON state, respectively. We denote the number of free mRNA molecules by $m$, the number of assembled mRNA molecules by $a$, and the total mRNA molecules by $n$. We provide a list of default parameters used throughout the simulations in \autoref{tab:parameters}.
\begin{table}[htbp]
\centering
\caption{Reaction list and corresponding propensities in the unified stochastic model.}
\label{table1}
\begin{threeparttable}
\setlength{\tabcolsep}{6pt}
\renewcommand{\arraystretch}{1.2}
\begin{tabularx}{\linewidth}{@{}c l r@{}}
\toprule
\# & Reaction & \text{Propensity} \\
\midrule
1 & Gene activation: $g=0 \to 1$ & $k_{\mathrm{on}}$ \\
2 & Gene inactivation: $g=1 \to 0$ & $k_{\mathrm{off}}$ \\
3 & Transcription: free mRNA production& $k_m = (\text{burst size}) \cdot k_{\mathrm{off}}$ \\
4 & Free mRNA degradation: $m \to m-1$ & $\gamma_m \cdot m$ \\
5 & assembled mRNA degradation: $a \to a-1$ & $\gamma_a \cdot a$ \\
6 & assembly: $m \to a$ & $\delta \cdot n \cdot m$ \\
7 & Disassembly: $a \to m$ & $\gamma \cdot a$ \\
\bottomrule
\end{tabularx}
\begin{tablenotes}[flushleft]
\footnotesize
\item[a] Reaction 3 fires only when the promoter is ON (\(g=1\)). Here \(n=m+a\).
\end{tablenotes}
\end{threeparttable}
\end{table}

\begin{table}[htbp]
\centering
\begin{threeparttable}
\caption{Default parameter values used in the unified stochastic model.}
\label{tab:parameters}
\begin{tabular}{lll}
\toprule
Parameter & Description & Default value \\
\midrule
$k_{\mathrm{on}}$  & Gene activation rate (burst frequency) & $0.05/600$ \\
$k_{\mathrm{off}}$ & Gene inactivation rate & $0.5/600$ \\
$k_m$              & Transcription rate (burst size $\times k_{\mathrm{off}}$) & $20 \times k_{\mathrm{off}}$ \\
$\delta$           & assembly rate constant & $0.05$ \\
$\gamma$           & Dissociation rate constant & $0.2$ \\
$\gamma_m$ & Free mRNA degradation rate & $0.05/60$ \\
$\gamma_a$  & assembled mRNA degradation rate & $0.05/60$ \\
$\langle n \rangle$ & average target mRNA number & $20$\\
\bottomrule
\end{tabular}
\begin{tablenotes}[flushleft]
\footnotesize
\item[a] Bursting parameters ($k_{on}$, $k_{off}$, $k_{m}$) are based on \cite{Rijntjes2020}.
\item[b] Decay parameters ($\gamma_m$, $\gamma_a$) are based on \cite{Munchel2011}.
\end{tablenotes}   
\end{threeparttable}
\end{table}
We first simulated mRNA assembly and disassembly inside a closed system with two molecular states, i.e., a conserved total number of mRNAs. Denote the number of free molecules by $m$, and the number of assembled molecules by $a$. Let $\delta$ be the assembled rate constant and $\gamma$ be the dissociation rate constant. Then the deterministic kinetics are
\begin{equation}
    \frac{da}{dt}=\delta n m - \gamma a,
    \label{eq: da/dt}
\end{equation}
\begin{equation}
   \frac{dm}{dt}=-\delta n m + \gamma a, 
\end{equation}
where $n=m+a$. The non-linear rate of aggregation $\delta n m $ may be expanded as 
\begin{equation}
    \delta n m =\delta m(m+a)=\delta m^2 +\delta a m,
\end{equation}
the term $\delta m^2$ can be interpreted as the rate at which two free molecules dimerize, and the term $\delta a m$ models assembly of free molecules with existing assemblies.
In steady state, if $n$ is held constant, then:
\begin{equation}
  m=\frac{\gamma n}{\delta n +\gamma}, 
  \label{eq: steady_state_m_n}
\end{equation}
and in the limit of large $n$, we expect $m$ to approach a constant,
\begin{equation}
    m\approx \frac{\gamma}{\delta}.
    \label{eq: m approximate}
\end{equation}
\\
We consider the variability of the free mRNA, measured by $CV^2$. In particular, 
\begin{equation}
    CV^2(m)=\frac{\operatorname{Var}(m)}{\langle m \rangle ^2}=\frac{\langle m^2 \rangle - \langle m \rangle ^2}{\langle m \rangle ^2}.
\end{equation}
We calculate the moments of a; $\langle a \rangle$ and for $\langle a^2 \rangle$ by using the chemical master equation, which can be shown that for any continuously differentiable function $\psi(\mathbf{N},t)$ \citep{hespanha2005model},

\begin{equation}
\label{moment_eq}
    \frac{d \langle \psi(\bm{N},t) \rangle}{dt} = \bigg \langle \sum_{i=1}^n \big[\psi(\bm{N}+\bm{\eta_i},t)-\psi(\bm{N},t)  \big] a_{i}(\bm{N}) \bigg \rangle,
\end{equation}
where $\bm{N}$ represents the full state of the system, i.e., $\bm{N}=(m,a)$; $\bm{\eta_i}$ is the state change vector for reaction $i$, e.g., $\bm{\eta}_{\mathrm{transcription}}=(1,0)$ for transcription $m\mapsto m+1$, $a_i(\bm{N})$ is the probability rate at which reaction $i$ occurs, given the current state $\bm{N}$.\\
In our context, we have
\begin{equation}
 \frac{d\langle a \rangle}{dt}=\langle \delta n(n-a)\rangle -\langle \gamma a\rangle=\delta n^2 -(\delta n + \gamma) \langle a \rangle.   
\end{equation}
\begin{equation}
   \frac{d\langle a^2 \rangle}{dt}= \delta n^2 + (2\delta n^2-\delta n +\gamma)\langle a \rangle-2(\delta n +\gamma)\langle a^2 \rangle.
\end{equation}
At steady state, $\dfrac{d\langle a \rangle}{dt}=\dfrac{d\langle a^2 \rangle}{dt}=0$, from which we deduce steady state $\langle \cdot \rangle_{\mathrm{ss}}$ values for the two moments.
\begin{equation}
\langle a \rangle_{\mathrm{ss}}=\frac{\delta n^2}{\delta n +\gamma},~ \langle m \rangle_{\mathrm{ss}}=n-\langle a \rangle_{\mathrm{ss}}=\frac{n\gamma}{\delta n +\gamma}. 
\end{equation}
\begin{equation}
    \langle a^2 \rangle_{\mathrm{ss}}=\dfrac{\delta n^2 + (2\delta n^2-\delta n +\gamma)\langle a \rangle_{\mathrm{ss}}}{2(\delta n +\gamma)}.
\end{equation}
The variance of $a$ at steady state gives:
\begin{equation}
 \operatorname{Var}(a)_{\mathrm{ss}}=\frac{\gamma \langle a \rangle_{\mathrm{ss}}}{\delta n +\gamma}=\frac{\delta n^2 \gamma}{(\delta n +\gamma)^2}.   
\end{equation}
Since $m=n-a$ where $n$ is a constant, $\operatorname{Var}(m)_{\mathrm{ss}}=\operatorname{Var}(a)_{\mathrm{ss}}$, and:
\begin{equation}
  CV^2(m)_{\mathrm{ss}}=\dfrac{\frac{\gamma \delta n^2}{(\delta n + \gamma)^2}}{\left(\frac{n\gamma}{\delta n +\gamma}\right)^2}=\dfrac{\delta}{\gamma}.
  \label{eq:cv2}
\end{equation}

\begin{figure}[htbp]
    \centering
    \includegraphics[width=\linewidth]{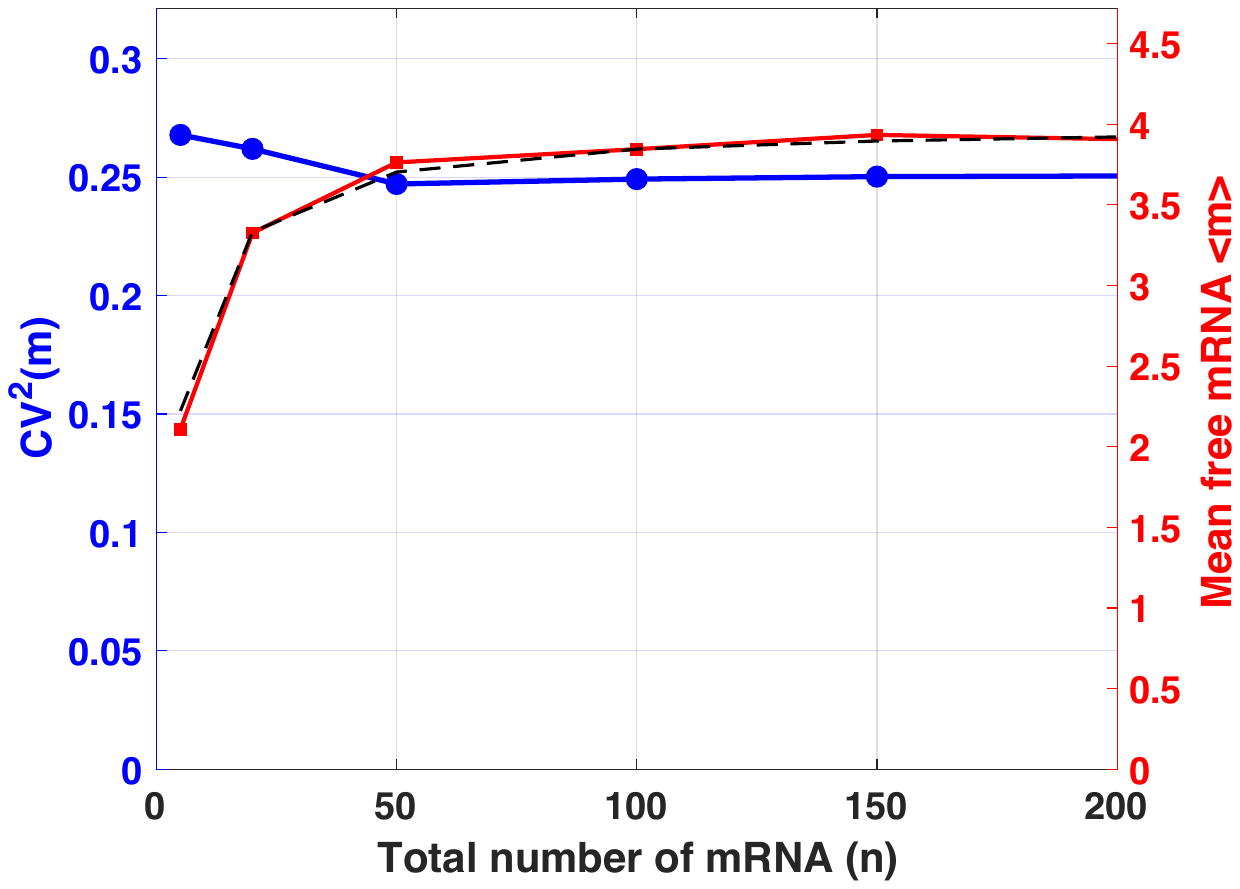}
    \caption{Dependence of $\langle m \rangle$ and $CV^2(m)$ on the total number of expressed mRNAs, $n$, in a closed system (constant $n$). Blue circles represent simulated noise $CV^2$ obtained from stochastic simulations. Red squares are the $\langle m \rangle$: simulated mean of free molecules, while the black dashed line is the theoretical prediction. The average number of free mRNA is maintained at around 4, and $CV^2(m)$ approaches a baseline as in \ref{eq:cv2}: $0.25=\delta/\gamma$, as $n$ increases. Simulation results under default parameters (\autoref{tab:parameters}).}
    \label{fig:2}
\end{figure}

Importantly, we predict that although the mean number of free mRNAs is dependent upon $n$, and reaches a limiting value as $n\rightarrow \infty$, the fluctuations in the available number are steady, independent of $n$, see \autoref{fig:2}.

Like in a phase separating model, assembly and disassembly maintain $m$ at a threshold abundance that becomes more and more insensitive to $n$ as $n$ increases. However, the variability in the mRNA abundance does not decrease as $n\rightarrow \infty$. By contrast, phase separation models assume a thermodynamic limit, meaning that relative fluctuations decrease as $n$ increases.
Our simulations expose the basic cost of mRNA assembly -  with a fixed total number, $n$, of mRNAs, the baseline variability in free mRNA becomes $\delta/\gamma$, no matter how large $n$ is. Hence, assembly can reduce mRNA variability only when this baseline variability is less than the variability in $n$.

Since we have assumed $n$ to be constant, the above variance is actually $\operatorname{Var}(m|n)$.
Using the law of total variance:
\(\operatorname{Var}(X)=\mathbb{E}(\operatorname{Var}(X|Y))+\operatorname{Var}(\mathbb{E}(X|Y))\), we obtain, if the pdf, $P(n)$, of $n$ is known:
\begin{equation}
\label{eq: variance}
\begin{aligned}
   \operatorname{Var}(m)
   &= \mathbb{E}\left(\frac{\delta n^2 \gamma}{(\delta n +\gamma)^2}\right) + \operatorname{Var}\left(\frac{n\gamma}{\delta n +\gamma} \right)\\
   &= \sum_n P(n) \left(\frac{\delta n^2 \gamma + n^2 \gamma^2}{(\delta n +\gamma)^2}  \right)- \left( \sum_n P(n)\left(\frac{n\gamma}{\delta n + \gamma}\right) \right)^2
\end{aligned}
\end{equation}
which allows us to calculate the $CV^2$ when $n$ is a random variable with any known (including empirical) distribution.

We first analyzed a system with a uniform decay rate, where the decay rates for assemblies and free mRNAs are the same. In this case, the simulation for the distribution of transcription is not affected by the downstream assembly and disassembly.
Then, we analyzed the case when the free mRNA degradation rate and the assembled mRNA degradation rate are different, specifically, we consider the cases when $\gamma_m=\gamma_a$, $\gamma_m \gg \gamma_a$, and $\gamma_a \gg \gamma_m$, with the last two limits leading respectively to the buffer model and filter model. In such coupled systems, the transcriptional output is influenced by the downstream processes, and numerical simulations provide a more accurate characterization of the resulting distributions.

\section*{Models for bursty gene transcription}
\subsection*{Noise analysis for Poisson distribution in transcription}
We first analyze $CV^2(m)$ in a counterfactual case of transcription being modeled as a Poisson process with mean $\lambda = \langle n \rangle $, then we calculated the pdf of $n$ analytically: 
\begin{equation}
    P(n)=\frac{\lambda^n e^{-\lambda}}{n!}, 
\end{equation}
which yields:
\begin{equation}
    \operatorname{Var}(n)=\langle n \rangle=\lambda, ~ CV^2(n)=\frac{1}{\langle n \rangle}.
\end{equation}
This representation requires two underlying Poisson-type assumptions: firstly, promoter activity switches on and off as a Poisson process, secondly, the events governing mRNA lifetimes (transcription and decay) are also Poisson processes \citep{Paulsson2005, Dattani2016}. Initially, we assume that the decay rate of mRNAs is independent of whether mRNAs are free or within RNPs. In the steady state, these assumptions yield a Poisson distribution, support for which has been found in certain housekeeping genes in yeast, and which is sometimes called the quiet limit of gene expression \citep{Zenklusen2008}.

\begin{figure}[htbp]
    \centering
    \includegraphics[width=\linewidth]{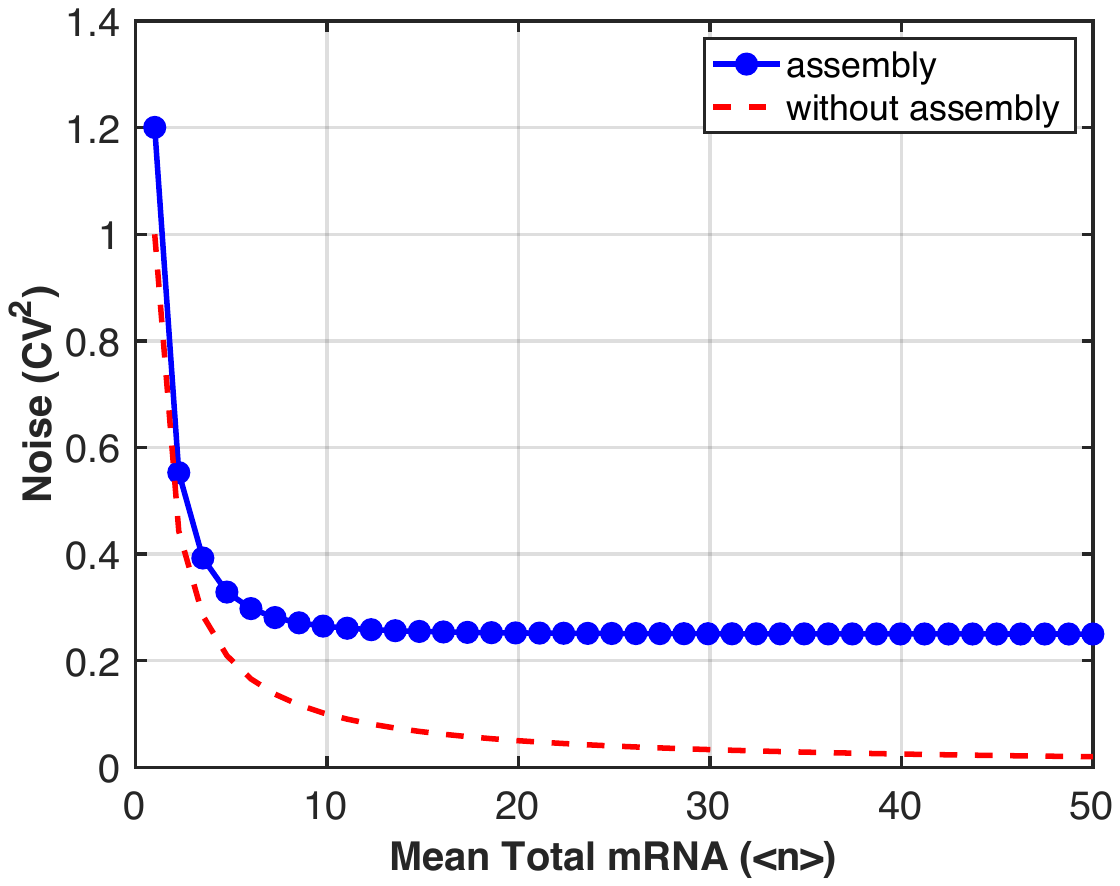}
    \caption{Comparing the noise ($CV^2$) in mRNA abundances under Poisson without (red dashed line) and with (blue circles) assembly and disassembly. As $\langle n \rangle$ increases, downstream noise approaches $\frac{\delta}{\gamma}=0.25$, whereas Poisson noise decreases to 0. Simulation results under default parameters (\autoref{tab:parameters}).}
    \label{fig:3}
\end{figure}

We compare the Poisson noise in $n$ with the downstream noise in $m$ predicted by our proposed model in \autoref{fig:3}. The Poisson noise is smaller than the downstream noise for all values of $\langle n \rangle$. This is easiest understood when $\langle n \rangle \rightarrow \infty$. In this limit, $CV^2(n)$ decays to $0$. However, the intrinsic noise of assembly and disassembly of RNPs imposes a floor of $\frac{\delta}{\gamma}=0.25$ on $m$. In our following analyses, we have the theoretical Poisson noise as a baseline for purely random transcription, and consider the effect of RNP assembly upon the much noisier process of bursty transcription.

\subsection*{Telegraph Model: analytic distribution and negative binomial limit}
\begin{figure}[htbp]
    \centering
    \includegraphics[width=\linewidth]{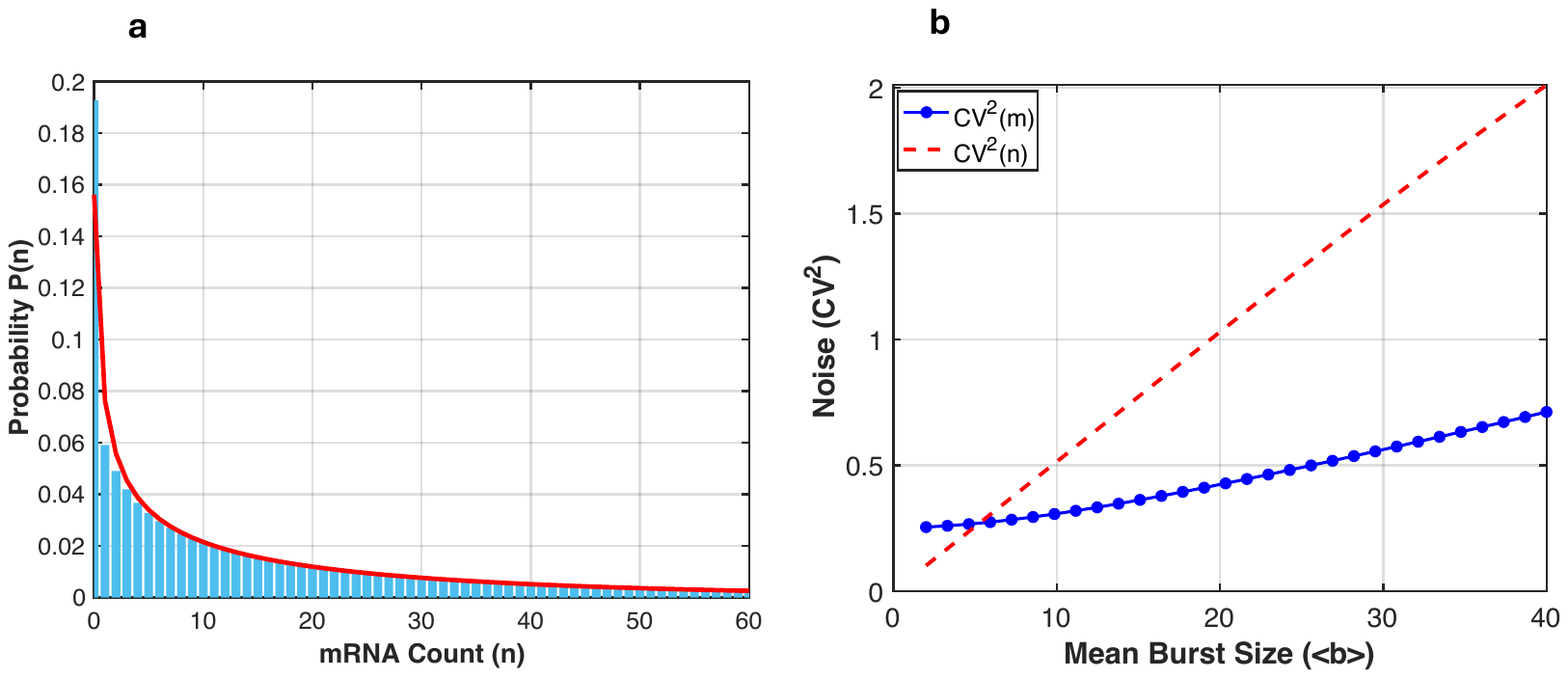}
    \caption{\textbf{(a)} Probability distribution of mRNA count $n$. Blue bars: the histogram represents the simulated probability $P(n)$ obtained from Gillespie simulations. Red curve: theoretical prediction.\\
    \textbf{(b)} Dependence of $CV^2(n)$ and $CV^2(m)$ as a function of mean transcriptional burst size $\langle b \rangle$. Blue circles: $CV^2(m)$ among assembly-forming mRNAs. Red dashes: $CV^2(n)$ for all transcribed mRNAs. As burst size increases, RNA-protein assembly becomes a noise-reducing mechanism. Simulation results under default parameters (\autoref{tab:parameters}).}
    \label{fig:4}
\end{figure}

In the Telegraph Model of gene expression, we simplify expression by a single gene into a random switch that has two states: the ON State, where it can actively transcribe and produce mRNA molecules, and the OFF State, during which it produces no mRNA \citep{boeger2015structural}. $k_{\mathrm{on}}$ describes the propensity for the gene to switch to the ON state, and $k_{\mathrm{off}}$ describes the propensity for the gene to switch back to the OFF state. The reciprocal $\dfrac{1}{k_{\mathrm{off}}}$ is the burst duration. The burst duration and the transcription rate while ON, denoted by $k_m$, together determine the mean burst size $\langle b \rangle$:
\begin{equation}
 \langle b\rangle =\frac{k_m}{k_{\mathrm{off}}}.   
\end{equation}
The steady-state distribution of mRNA counts is given by a Poisson-Beta mixture \citep{Kim2013, Cavallaro2021}:
\begin{equation}
    P(n)=\int_0^1 \frac{e^{-\lambda x}(\lambda x)^n}{n!}\cdot \frac{x^{\alpha -1}(1-\alpha)^{\beta -1}}{B(\alpha, \beta)}dx, ~n=0,1,2,\ldots
\end{equation}
with parameters
\begin{equation}
    \alpha = \frac{k_{\mathrm{on}}}{\gamma}, ~\beta= \frac{k_{\mathrm{off}}}{\gamma},~ \lambda=\frac{k_m}{\gamma}.
\end{equation}
Moveover,
\begin{equation}
 \langle n \rangle = \lambda \frac{\alpha}{\alpha+\beta},~ \operatorname{Var}(n)=\langle n \rangle+\lambda^2 \frac{\alpha \beta}{(\alpha+\beta)^2(\alpha+\beta+1)}.   
\end{equation}
Thus, the Fano factor \citep{Grima2024_BiophysJ} is 
\begin{equation}
    F=\frac{\mathrm{Var}(n)}{\langle n\rangle} =1 +\lambda \frac{ \beta}{(\alpha+\beta)(\alpha+\beta+1)}.
\end{equation}
In the bursty regime of the two-state telegraph model ($\beta \gg 1$), as noted in \citep{Grima2024_BiophysJ, ShahrezaeiSwain2008_PNAS}, the Fano factor reduces to 
\begin{equation}
F \rightarrow 1 + \langle b \rangle.
\end{equation}
Moreover, the distribution converges to a negative binomial distribution (shown in \autoref{fig:4} \textbf{(a)}) \citep{Cao2020_PNAS, Jia2023}:
\begin{equation}
    P(n)= {{n+r-1} \choose n}p^r (1-p)^n
\end{equation}
with $r=k_{\mathrm{on}}/\gamma$, $p=1/(1+\langle b \rangle)$, enabling us to calculate $CV^2(m)$ from \autoref{eq:cv2}.
For comparison, we calculate $CV^2(n)$:
\begin{equation}
   CV^2(n)=\frac{F}{\langle n \rangle} =\frac{1+\langle b \rangle}{\langle n \rangle}.
   \label{eq:fano-cv2}
\end{equation}
If $\langle b \rangle$ is positive, the Fano factor is always greater than 1, confirming that bursty transcription is noisier than a simple Poisson process.

Noise continues to increase if $\langle n \rangle$ is held constant but $\langle b \rangle$ is increased, meaning that the nucleus produces mRNAs in more intense but less frequent bursts. On the other hand, for a fixed mean burst size $\langle b \rangle$, the noise is inversely proportional to the mean number of molecules $\langle n \rangle$, meaning that more abundant mRNAs tend to have lower noise.

When we add assembly and disassembly to the model, continuing to assume that mRNAs have the same decay rate whether they are free or in assemblies, we find that the growth of $CV^2(n)$ with $\langle b \rangle$ means that even for modest burst intensities, assembly of RNAs into RNPs leads to less copy number variability in free mRNAs than in expressed mRNAs (\autoref{fig:4} \textbf{(b)}). Since $\langle m \rangle$ increases only slowly with $\langle b \rangle$, we can estimate the threshold value of $\langle b \rangle$ at which assembly becomes an effective noise reduction mechanism:
\begin{equation}
    \frac{1+\langle b \rangle}{\langle n \rangle}=\frac{\delta}{\gamma}.
\end{equation}
For our default parameters, the estimated threshold is: 
\begin{equation}
    \langle b \rangle = \frac{\delta}{\gamma}\langle n \rangle -1 = 4,
\end{equation}
in close accord with simulations.
\section*{Analysis for different decay rates}
Hitherto, we have assumed that assembly and disassembly have no effect upon the lifetimes of mRNAs. In other words, the same degradation rate can be applied to free mRNAs as to the assembled population. In real cells, associations with proteins may protect RNAs from the proteins responsible for degradation as well as from transcription. Or conversely, they may hasten the degradation of mRNAs.
We consider both of these scenarios. Since analytic results are no longer possible for these systems, we study the variability of mRNA copy numbers by stochastic simulations. Specifically, we simulate all 7 processes diagrammed in \autoref{table1} using Gillespie's algorithm. Assembly and disassembly rates are given in \autoref{tab:parameters}.

 \subsection*{Buffer model}
In the buffer model, we consider the limit when only free mRNAs decay, i.e., $\gamma_a = 0$, $\gamma_m = 0.05/60$. We consider two cases: when the transcriptional burst size is large, i.e., $\langle b \rangle = 20$, $k_{\mathrm{on}}=0.05/600 s^{-1}$, and $k_{\mathrm{off}}=0.5/600 s^{-1}$, and when the burst size is small, i.e., $\langle b \rangle = 2.4$, $k_{\mathrm{on}}=0.05/60 s^{-1}$, and $k_{\mathrm{off}}=2.5/60 s^{-1}$, in both cases, assembly formation keeps the number of free mRNA small, even as the total copy of mRNA increases. We compare the downstream noise $CV^2(m)$ with upstream noise $CV^2(n)$ and with the Poisson Noise Floor, $\frac{1}{\langle n \rangle}$, by keeping $\gamma_m$ constant and
varying $\gamma_a$, using a long simulation time $1\times 10^6$ s to build up stochastic information about the occupation frequencies of each state.

To clarify the mechanism of noise reduction, we distinguish between ``upstream'' noise, which originates from bursty transcription and is reflected in the total mRNA $(n)$ fluctuations, and ``downstream'' noise, which is the buffered fluctuation experienced by the free mRNA $(m)$ available for translation.
\begin{figure}[htbp]
    \centering
    \includegraphics[width=\linewidth]{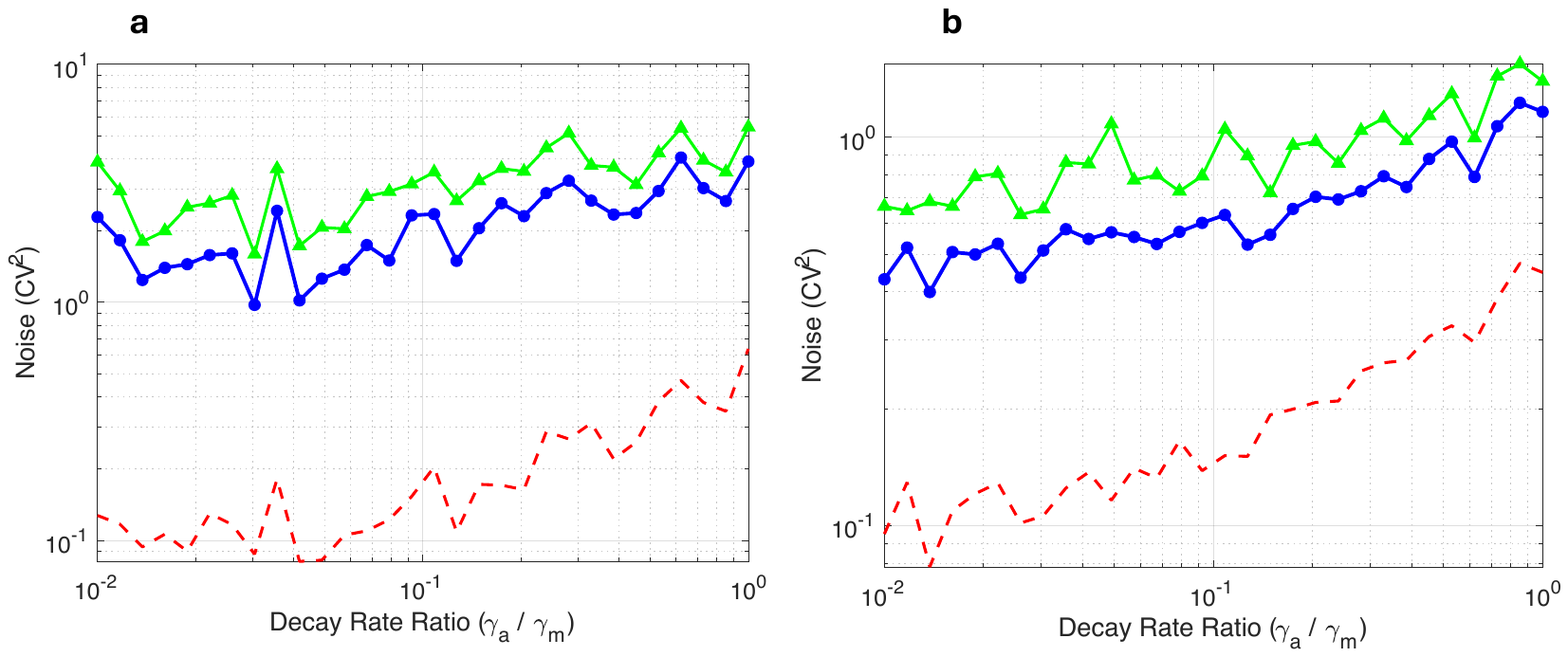}
    \caption{Noise measures: downstream noise $CV^2(m)$ (blue circles), upstream noise $CV^2(n)$ (green triangles), and the Poisson baseline ${1}/{\langle n \rangle}$ (red dashed) under buffering model with mean burst size \textbf{(a)} $\langle b \rangle=20$,
    \textbf{(b)} $\langle b \rangle=2.4$.\\
    In \textbf{(a)}, when $\gamma_a/\gamma_m=1$, $CV^2(m)/CV^2(n)=0.7173$;\\
    when $\gamma_a/\gamma_m=0.01$, ${CV^2(m)}/CV^2(n)=0.5879$.\\
    In \textbf{(b)}, when ${\gamma_a}/{\gamma_m}=1$, ${CV^2(m)}/{CV^2(n)}=0.8335$;\\
    when ${\gamma_a}/{\gamma_m}=0.01$, ${CV^2(m)}/{CV^2(n)}=0.663$.}
    \label{fig:5}
\end{figure}

\begin{figure}[htbp]
    \centering
    \includegraphics[width=\linewidth]{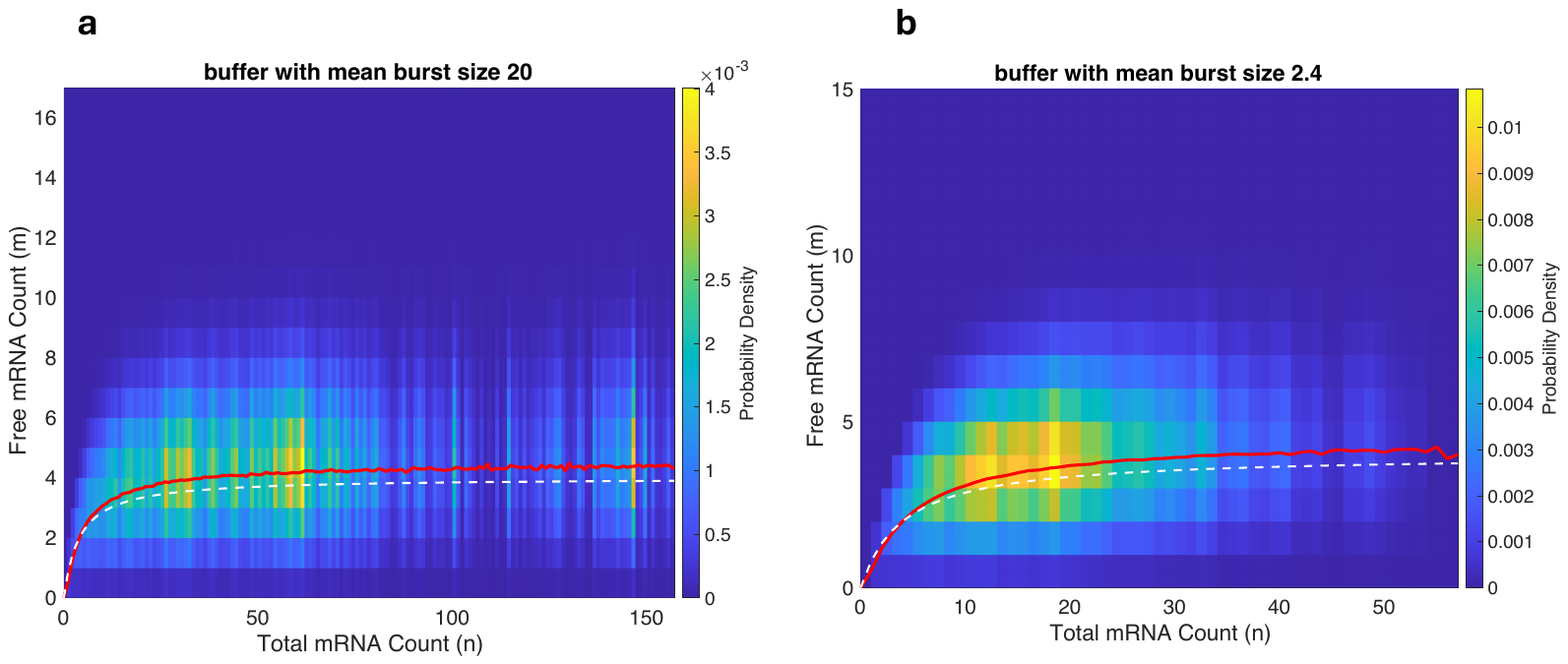}
    \caption{
    Probability density of free mRNA molecules, $m$, and total mRNA, $n$. Colors represent the frequency of simulation outcomes, with warmer colors indicating a higher frequency.\\
    The white dashed curves are the theoretical prediction of free mRNA number based on rapid kinetics of assembly and disassembly, and quasi steady $n$, and the red curves are the simulated free mRNA number.\\
    \textbf{(a)} parameters $\gamma_a = 0$, $\gamma_m = 0.05/60$, $k_{\mathrm{on}}=0.05/600 s^{-1}$, $k_{\mathrm{off}}=0.5/600 s^{-1}$, with mean burst size $\langle b \rangle = 20$. \\
    \textbf{(b)} parameters $\gamma_a = 0$, $\gamma_m = 0.05/60$, $k_{\mathrm{on}}=0.05/60 s^{-1}$, $k_{\mathrm{off}}=2.5/60 s^{-1}$, with mean burst size $\langle b \rangle = 2.4$.
    }
    \label{fig:6}
\end{figure}

We observed from \autoref{fig:5} that whether $\langle b \rangle = 2.4$ or $\langle b \rangle = 20$, the downstream noise is lower than the upstream noise, indicating that the buffer model can mitigate the noise produced by intense and even moderate amounts of bursting. Plotting system states in terms of the number of free ($m$) and total ($n$) mRNAs (see \autoref{fig:6}) clarifies the mechanism of noise reduction by long-lived mRNA-protein assemblies. Newly produced mRNAs are quickly absorbed into the assemblies. Moreover, the mRNAs in the assemblies are slowly depleted over time since they need to re-enter the dilute phase. We emphasize, though, that buffering through assembly doesn't achieve variabilities comparable to the Poisson noise floor.
 
\subsection*{Filter Model}
Under alternative circumstances, interactions with RNPs can destabilize mRNA by modifying the transcript and targeting it for degradation. The RNP droplets may act as a filter, causing mRNAs within the cytosol to be quickly degraded and recycled. In this case, we assume that $\gamma_a \gg \gamma_m$ in \autoref{table1}.

\begin{figure}[htbp]
    \centering
    \includegraphics[width=\linewidth]{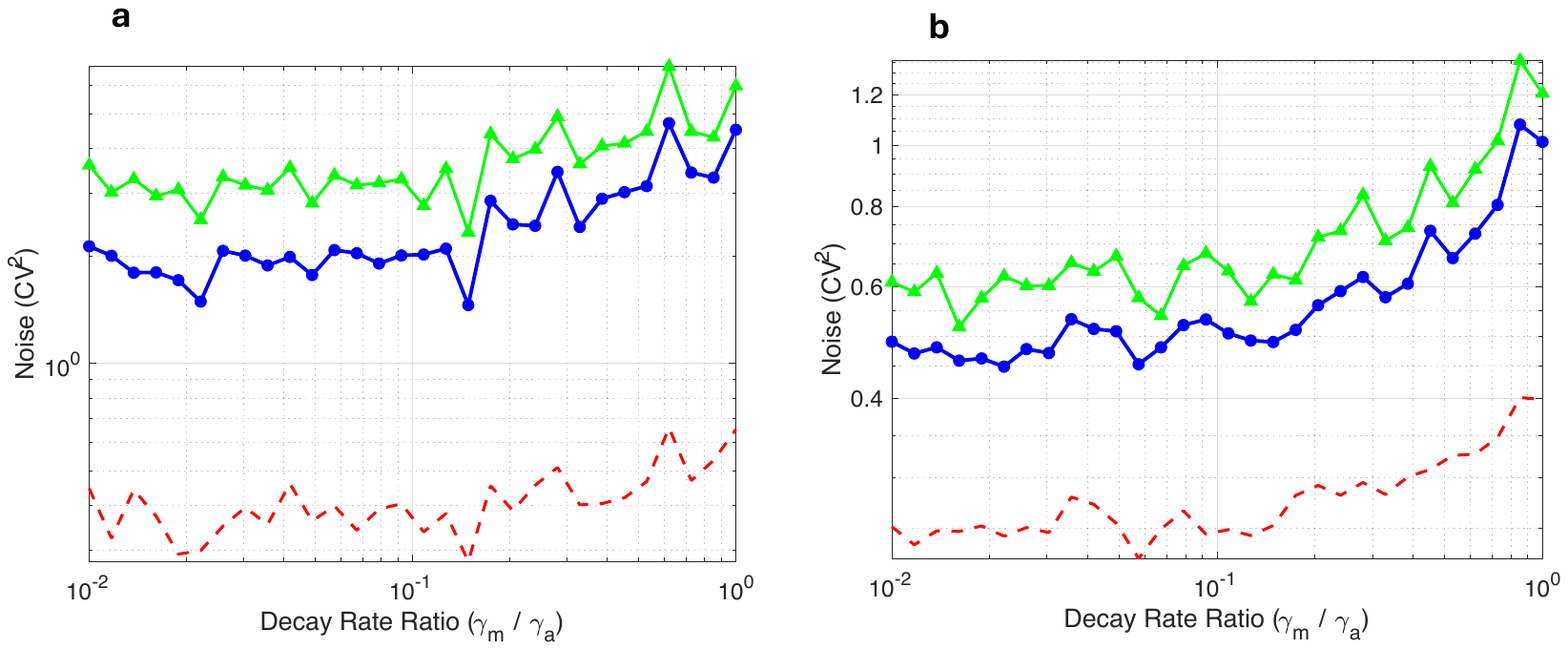}
    \caption{Noise measures: downstream noise $CV^2(m)$ (blue circles), upstream noise $CV^2(n)$ (green triangles), and the Poisson baseline ${1}/{\langle n \rangle}$ (red dashed) under filtering model with mean burst size \textbf{(a)} $\langle b \rangle=20$,
    \textbf{(b)} $\langle b \rangle=2.4$.\\
    In \textbf{(a)}, when ${\gamma_a}/{\gamma_m}=1$, ${CV^2(m)}/{CV^2(n)}=0.7549$;\\
    when ${\gamma_a}/{\gamma_m}=100$, ${CV^2(m)}/{CV^2(n)}=0.5933$.\\
    In \textbf{(b)}, when ${\gamma_a}/{\gamma_m}=1$, ${CV^2(m)}/{CV^2(n)}=0.8398$;\\
    when ${\gamma_a}/{\gamma_m}=100$, ${CV^2(m)}/{CV^2(n)}=0.8062$.}
    \label{fig:7}
\end{figure}

\begin{figure}[htbp]
    \centering
    \includegraphics[width=\linewidth]{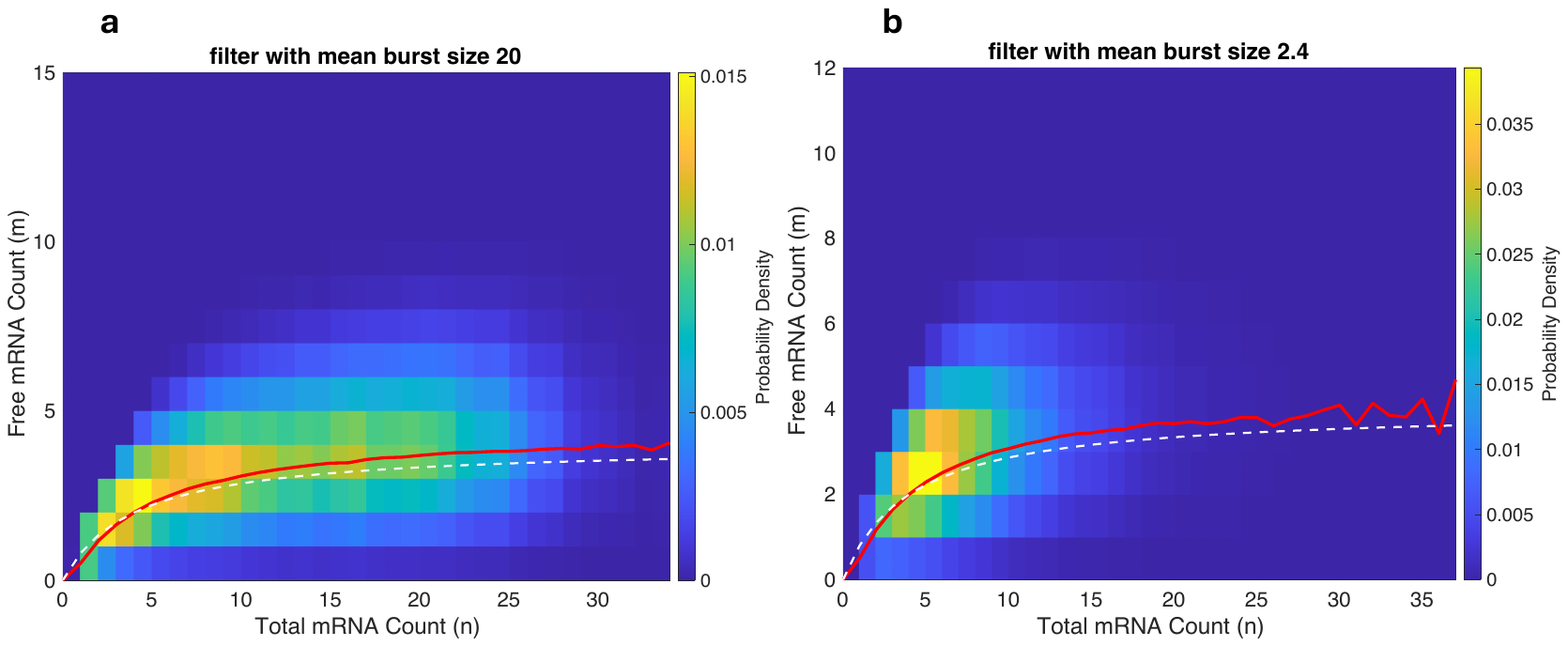}
    \caption{Probability density of free mRNA molecules, $m$, and total mRNA, $n$. Colors represent the frequency of simulation outcomes, with warmer colors indicating a higher frequency.\\
    The white dashed curves are the theoretical prediction of free mRNA number based on rapid kinetics of assembly and disassembly, and quasi steady $n$, and the red curves are the simulated free mRNA number.\\
    \textbf{(a)} parameters $\gamma_a = 0.05/60$, $\gamma_m=0.001/60$, $k_{\mathrm{on}}=0.05/600 s^{-1}$, $k_{\mathrm{off}}=0.5/600 s^{-1}$, with mean burst size $\langle b \rangle = 20$. \\
    \textbf{(b)} parameters $\gamma_a = 0.05/60$, $\gamma_m=0.001/60$, $k_{\mathrm{on}}=0.05/60 s^{-1}$, $k_{\mathrm{off}}=2.5/60 s^{-1}$, with mean burst size $\langle b \rangle = 2.4$.}
    \label{fig:8}
\end{figure}

Again, we perform numerical simulations of the complete system (\autoref{table1}) with two different burst intensities $\langle b \rangle=2.4$ and $\langle b \rangle=20$. We find that at both intensities, there is significant noise reduction compared to the system without assembly, though the degree of noise reduction ($41\%$ versus $19\%$) is systematically greater for larger burst intensities (\autoref{fig:7}). Again, we gain mechanistic insight into how assemblies function to reduce copy number noise by studying the joint probability density function of $m$ and $n$ (\autoref{fig:8}). For both burst intensity means, we find that the mean number of free mRNAs closely matches as a function of $n$ to \autoref{eq: steady_state_m_n}, indicating that rapid mRNA assembly and disassembly keep free and assembled mRNA populations in a quasi-equilibrium. However, rapid depletion of the mRNAs pushes the system toward smaller overall mRNA abundances (compare \autoref{fig:8} \textbf{(a)} with \autoref{fig:6} \textbf{(a)}, for example), where noise is greater.

\section*{Comparison with experimental data}
\subsection*{Analysis of transcript abundances for an experimental RNP-forming system}

\begin{figure}[htbp]
    \centering
    \includegraphics[width=\linewidth]{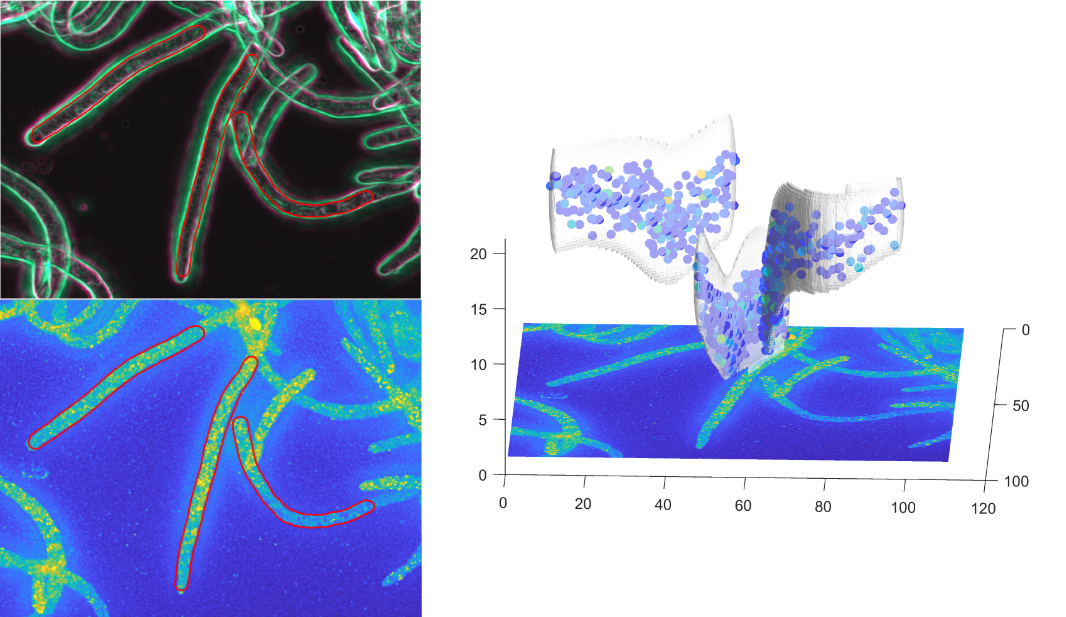}
    \caption{3D segmentation of fungal cells in \textit{Ashbya gossypii}. Top left panel: maximum intensity projection of a stack of phase contrast images of the cells, processed using steerable filters to highlight cell boundaries. Images are false colored so that edges in the middle of the $z$-stack are colored green, and edges within 3$\mu$m of the top or bottom layer are colored magenta, allowing cells that are not completely contained in the $z$-stack to be ignored. Three hyphae are shown outlined. Bottom left panel: Fluorescence image showing the \textit{CLN3} mRNA spots. These spots are color-coded by the inferred number of mRNA molecules they contain, ranging from 1 (blue) to 12 (yellow). Right panel: 3D reconstruction of the 3D cell surfaces, suspended above this fluorescence image.}
    \label{fig:9}
\end{figure}
We tested whether our modeling could be consistent with previously collected data on RNP-forming systems. Distributions of \textit{CLN3} mRNA transcripts -- which are translated into the cyclin Cln3 -- were mapped in the filamentous fungus \textit{Ashbya gossypii} (see \autoref{fig:9}) using single-molecule fluorescence in-situ hybridization (smFISH) \citep{lee2013protein}. \textit{CLN3} was chosen because it is known to form RNP droplets with Whi3 protein, and the mechanism of \textit{CLN3}-Whi3 RNP assembly has been related to the presence of Whi3 binding sequences within the mRNA, and to polyQ-tract driven assembly of Whi3 proteins \citep{zhang2015rna}.
Based on prior results about spheres of mRNA enrichment surrounding nuclei  \citep{dundon2016clustered}, we partitioned the cytoplasm (i.e., 3D segmented hyphae, subtracting their nuclei) into spheres of radius 2.5 $\mu$m centered at nuclei centroids (the average nucleus radius $\approx 1.9 \mu$m). We excluded the nucleus itself and the parts of the sphere that extended outside of the hypha. Within each sphere, we measured the mRNA concentration (total number of mRNA transcripts divided by neighborhood volume), as well as the mean number of mRNAs per spot. Spots with weight $> 1$ were considered to be in a condensate. We also detected mRNA transcripts within nuclei. A nucleus was assumed to be actively transcribing mRNAs if it contained mRNA transcripts.

\begin{figure}[!htb]%
    \centering
    \subfloat[]{{\includegraphics[width=.45\linewidth]{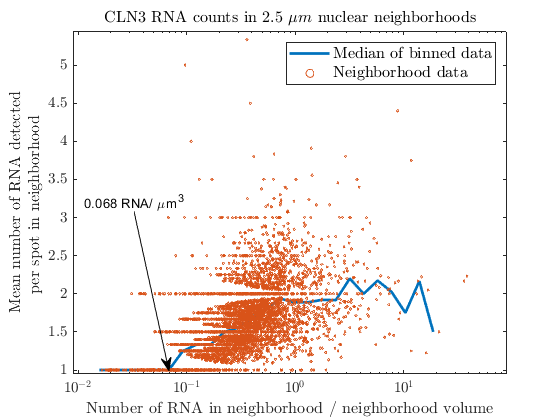} }}%
    \qquad
    \subfloat[]{{\includegraphics[width=.45\linewidth]{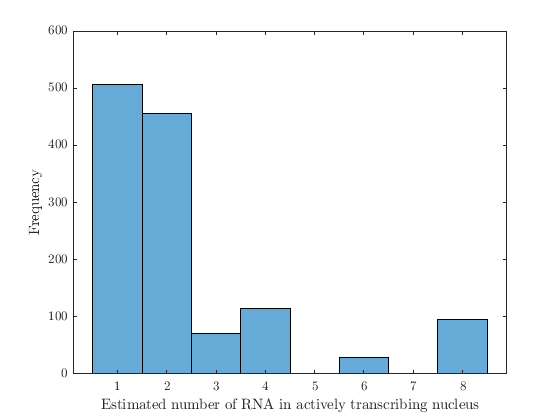} }}%
    \caption{Spheres of radius $2.5 \mu m$ were constructed around each nucleus to carve out cytoplasmic regions we refer to as \textit{nuclear neighborhoods}. Left panel: In each nuclear neighborhood, we measured the number of mRNAs divided by the neighborhood volume, and the mean spot weight. From this we formed a scatter plot, and binned the data in $40$ uniform compartments. In each bin, we measured the median mean spot weight to form the curve in blue. This curve experiences a transition at a concentration of $\approx 0.068$ mRNA,$/\mu m^3$ signifying the minimum assembly size. With a mean neighborhood volume of $\approx 37 \mu m^3$, we estimate that assembly holds free mRNA at average number of $\approx 2.5$ free mRNAs per neighborhood. Right panel: A histogram of the number of mRNA in actively transcribing nuclei, which we use to approximate the distribution of bursts. The mean number of transcripts synthesized in a burst was measured to be $\approx 2.4$. }%
    \label{fig:10}%
\end{figure}
In Fig. \ref{fig:10}a, binning of nuclear neighborhood data reveals evidence of an assembly threshold at a concentration of $\approx$ 0.068 mRNA$/\mu m^3$. The volume of a typical cytoplasmic neighborhood is $\approx 37 \mu m^3$, placing the assembly threshold at $\approx 2.5$ total mRNA transcripts per neighborhood. This number matches very closely to the mean number of mRNA transcripts that were detected in actively transcribing nuclei (2.4), which we interpret as the burst size parameter (see Fig. \ref{fig:10}b). \textit{CLN3} may be only one of multiple RNAs contributing to the formation of mRNA-protein assemblies, but the data are consistent with \textit{CLN3} being the primary species of mRNA controlling assembly formation.

\subsection*{Comparison with \textit{CLN3} data}
The biological data on the \textit{CLN3} model RNP-forming system analyzed above were broadly supportive of droplets playing a role in suppressing free mRNA fluctuations. Specifically, we found that the threshold abundance at which assembly was evidenced to occur ($\approx$2.5 per nuclear neighborhood) was broadly comparable to the burst size inferred by counting mRNAs within nuclei (2.4). This numerical agreement suggests that a typical burst is just sufficient to trigger local aggregation. Although this does not provide direct evidence of noise suppression, the match between burst size and condensation threshold supports the plausibility of the proposed mechanism. The relationship between optimal burst size and assembly threshold is not deeply dependent upon whether condensates are filters or buffers, or equivalently, whether sequestration within a condensate increases or decreases mRNA lifetimes. Indeed, available experimental evidence suggests that either may be possible, including in homologous systems. For example, in data from yeast cells, which contain near-homologs of \textit{CLN3} and Whi3, suppressing Whi3 protein expression extends the lifetimes of \textit{CLN3} mRNAs, suggesting that condensates destabilize mRNAs \citep{cai2013effects}. But data in the filamentous fungus \textit{A. gossypii} indicate that RNP droplets extend the lifetime of \textit{CLN3} mRNAs. The question of whether the droplets are functioning as buffers or as filters remains unanswered here, though the generality of our analysis supports noise reduction by either mechanism.

We include an additional two notes of caution here -- it is not yet possible to quantitatively measure the abundance of Cln3 proteins in \textit{Ashbya} at the scale of individual cells, so we do not offer direct proof that Cln3 protein noise is effected by RNP formation for this system. Additionally, the main assumption of our model: that mRNAs within droplets are largely translation-inert, remains a simplification. Recent work by \cite{Geisterfer2024}  indicates that translation of associated transcripts, including \textit{CLN3}, can be modulated by assembly state and is often repressed, though not universally abolished. Moreover, RNPs are assuming an ever more central role in cell biology, and there has been a corresponding expansion, in recent years, of techniques for perturbing and measuring RNP formation kinetics in live cells \citep{langdon2018mrna}, and we expect the predictions from our mathematical model to shortly become experimentally testable.

\section*{Conclusion}
In this work, we developed a stochastic framework to understand how mRNA assembly reduces gene expression noise. Unlike previous studies of phase separation, where mRNA abundances are large, our work focuses on the case where the noise can also be reduced when the system has a small number of mRNA transcripts. We investigated different models for bursty transcription and modeled noise suppression in different cases. We illustrated our model of assembly and disassembly first in a closed system, and analyzed the cost of noise and noise reduction. We coupled all stochastic models and linked the downstream assembly model with the upstream process of gene expression. We varied the decay rate when mRNA decay primarily occurs in the cytoplasm or occurs in the assemblies, in order to demonstrate that the noise-reducing ability is robust in both cases.

This study underscores the significance of complex formation in biological regulation as a dynamic noise-control strategy, even when the average number of mRNA is small. These insights suggest a mechanism in which cells can change the $\gamma_a / \gamma_m$ ratio to switch between noise-buffering and noise-filtering. This opens up opportunities for experimental investigation into how cells use this tuning to optimize precision.

Furthermore, our structure motivates the need for more experimental data, particularly in regimes of low mRNA abundance. Such measurements will be significant to validate the predictions and give powerful insight into parameter tuning in order to control cellular noise.

Funding support from the National Science Foundation (RoL-1840273) is gratefully acknowledged. XL was supported by an Overseas Research Fellowship from the University of Hong Kong. We thank Jiayue Liu for an early analysis of \textit{CLN3} abundances and Sierra Cole for experimental assistance.

\bibliographystyle{elsarticle-harv} 
\bibliography{main.bib}



\end{document}